\documentclass[twocolumn, pra, amssymb,superscriptaddress, aps,preprintnumbers,amsmath,floatfix]{revtex4} 
\usepackage{amsmath}
\usepackage{amssymb}
\usepackage{color,soul}
\usepackage{graphicx}  
\usepackage{mathrsfs}
\usepackage{datetime}
\usepackage[hidelinks]{hyperref}
\usepackage{multirow}
\usepackage{isotope}

\usepackage[dvipsnames]{xcolor}

\usepackage{pdfpages}

\def\Neff{N_{\rm eff}}
\def\NTh{N_{\rm Th}}

\begin{document}

\title{Driven electronic bridge processes via defect states in $^{229}$Th-doped crystals}

\author{Brenden S. Nickerson}
\email[]{brenden.nickerson@mpi-hd.mpg.de}
\affiliation{Max-Planck-Institut f\"ur Kernphysik, D-69117 Heidelberg, Germany}

\author{Martin Pimon}
\affiliation{Center for Computational Material Science, Technische Universit\"at Wien, 1040 Vienna, Austria}

\author{Pavlo V. Bilous}
\affiliation{Max-Planck-Institut f\"ur die Physik des Lichts, D-91058 Erlangen, Germany}
\affiliation{Max-Planck-Institut f\"ur Kernphysik, D-69117 Heidelberg, Germany}

\author{Johannes Gugler}
\affiliation{Center for Computational Material Science, Technische Universit\"at Wien, 1040 Vienna, Austria}

\author{Georgy A. Kazakov}

\author{Tomas Sikorsky}

\author{Kjeld Beeks}

\affiliation{Atominstitut, Technische Universit\"at Wien, 1020 Vienna, Austria}

\author{Andreas Gr\"uneis}
\affiliation{Institute for Theoretical Physics, Technische Universit\"at Wien, 1040 Vienna, Austria}
\affiliation{Center for Computational Material Science, Technische Universit\"at Wien, 1040 Vienna, Austria}

\author{Thorsten Schumm}
\affiliation{Atominstitut, Technische Universit\"at Wien, 1020 Vienna, Austria}

\author{Adriana P\'alffy}
\email[]{adriana.palffy-buss@fau.de}
\affiliation{Department of Physics, Friedrich-Alexander-Universit\"at Erlangen-N\"urnberg, D-91058 Erlangen, Germany}
\affiliation{Max-Planck-Institut f\"ur Kernphysik, D-69117 Heidelberg, Germany}

\date{\today}

\begin{abstract}
The electronic defect states resulting from doping $^{229}$Th in CaF$_2$ offer a unique opportunity to excite the nuclear isomeric state $^{229m}$Th at approximately 8 eV via electronic bridge mechanisms. We consider bridge schemes involving stimulated emission and absorption using an optical laser. The role of different multipole contributions, both for the emitted or absorbed photon and nuclear transition, to the total bridge rates are investigated theoretically. We show that  the electric dipole component is dominant for the electronic bridge photon. In contradistinction, the electric quadrupole channel of the $^{229}$Th isomeric transition plays the dominant role for the bridge processes presented. The driven bridge rates are discussed in the context of background signals in the crystal environment and of implementation methods. We show that inverse electronic bridge processes quenching the isomeric state population can improve the performance of  a solid-state nuclear clock based on $^{229m}$Th.
\end{abstract}
\maketitle

\section{Introduction}

The nuclear isomer $^{229m}$Th is our most compelling candidate for the development of the first nuclear clock. With an energy of just 8~eV \cite{B.Seiferle2019,Sikorsky2020}, it is more comparable to transitions of valence electrons in the atomic shell than anything expected in all of the currently known isotopes \cite{Wense_Nature_2016}. Most importantly, the $^{229m}$Th isomer could be accesible by narrow-band vacuum ultraviolet (VUV) lasers, which is the key to designing a frequency standard based on a nuclear transition \cite{Peik_Clock_2003,QST-Review2021}. A practical implementation will require development of such lasers and a more precise knowledge of the isomer energy. At present, the isomer energy was reported as $E_m=8.28(17)$~eV using a direct measurement of internal conversion electrons \cite{B.Seiferle2019},  $E_{m}=8.30(92)$ eV \cite{Yamaguchi_EnTh229m_2019} from determining the transition rates and energies from the above level at 29.2 keV in a calorimetric experiment, or $E_m=8.10(17)$ eV from state-of-the-art gamma spectroscopy measurements using a dedicated cryogenic magnetic microcalorimeter \cite{Sikorsky2020}.

Substantial experimental progress has been made in the study of thorium ions in beams and traps, with the first direct proof of isomer decay \cite{Wense_Nature_2016, Seiferle_PRL_2017}, an updated energy determination \cite{B.Seiferle2019} and the measurement of isomer nuclear moments \cite{Thielking2018}. A solid-state thorium oxide target has also been studied recently with x-ray nuclear resonance scattering pumping schemes for improved isomer production \cite{Masuda_Nature_2019}. Here we are interested in an alternate experimental approach making use of VUV-transparent crystals doped with thorium ions. The crystal environment allows for dopant densities many orders of magnitude larger than would be possible for trapped ions \cite{Kazakov_2012, Stellmer2018, Campbell2011, coulomb_crystal}. Concentrations in the range of $10^{16}-10^{18}$ cm$^{-3}$ are easily reached \cite{Capelli2015}, which make a significant impact on the stability of the potential clock proportionally to $\sqrt{N}$ \cite{PhysRevA.47.3554}, where $N$ is the number of interrogated nuclei. Along with the relative ease with which the doped crystals can be manufactured and transported, this makes thorium-doped VUV transparent crystals a promising candidate for the nuclear clock implementation.   

Despite the apparent upsides, significant effort has gone into attempts of direct isomer excitation within the VUV-transparent crystal environment so far without success \cite{Jeet_PRL_2015, Rellergert_test_2010, Stellmer2018, Stellmer2015, Dessovic_2014, crystaldamage, Zimmermann_thesis_2010}. Allegedly, theoretical models show that the radiative transition is weak \cite{Minkov_Palffy_PRL_2017,Minkov_Palffy_PRL_2019,Minkov_Palffy_PRC_2021}, and also the explored energy range
around the previously used energy value of 7.8 eV \cite{Beck_78eV_2007} might have been disadvantageous.
 In addition, a variety of crystal defects induced by radioactivity and laser irradiation led to reported background in the UV and VUV range along with a reduction in VUV transmission. Background sources include phosphorescence of crystal defects both intrinsic and laser-induced, and Cherenkov radiation stemming from $\beta$-radioactive daughter nuclei in the $^{229}$Th decay chain \cite{Rellergert_test_2010,Stellmer2015,Stellmer2018,Dessovic_2014, crystaldamage, Zimmermann_thesis_2010}.

Here we outline excitation methods that make use of a specific set of electronic defect states in the crystal to increase both the rate of excitation and the total excited population of the nuclear isomeric state. These defect states are predicted by density functional theory (DFT) to exist in the vicinity of the $^{229}$Th nucleus as a direct consequence of the crystal doping. Their energies lie in the band gap of CaF$_2$ close to the nuclear transition energy \cite{Dessovic_2014}. In Ref.~\cite{Nickerson20PRL} we have put forward how these states can be used to drive an electronic bridge (EB) scheme for excitation of the isomer in the crystal environment. The EB process can enable nuclear excitation and decay via electromagnetic coupling to the atomic shell in a third-order perturbation theory process, without requiring a perfect energetic match between the atomic and nuclear transitions. The energy mismatch is covered by the emission or absorption of a photon. In the context of $^{229}$Th, several EB scenarios  for Th ions have been investigated theoretically \cite{StrizhovTkalya_JETP_1991,TkalyaBridge1992,TkalyaBridge1992-2, PorsevFlambaum_Brige_PRL_2010,PorsevFlambaum_Brige1+_PRA_2010,PorsevFlambaum_Brige3+_PRA_2010, Bilous2018}. 

In this work we build up on the original proposal \cite{Nickerson20PRL} with a twofold purpose. First, we further investigate the role of different multipolarities, both for the emitted or absorbed photon (referred to here in general as the bridge photon) and the nuclear transition itself. In Ref.~\cite{Nickerson20PRL} we focused on EB processes where the optical bridge photon had electric dipole ($E1$) multipolarity which was assumed to be the dominant channel. To have a better understanding of the competing processes, here we analyze EB rates where the bridge photon has $E1$, magnetic dipole ($M1$) or electric quadrupole ($E2$) multipolarity respectively. Since the crystal wave functions are not eigenfunctions of angular momentum and parity, one cannot rule out {\it a priori} the effect of the $M1$ and $E2$ multipole operators. Nevertheless, these processes are shown to be orders of magnitude slower than the corresponding $E1$ process and therefore negligible here. Details regarding the density functional calculations which are crucial to the results presented here are also covered. 

The convergence criteria for the EB rates are studied and broken down into contributions from $M1$ and $E2$ nuclear transition multipolarities, respectively. Traditionally, earlier discussions of the potential decay pathways for the nuclear isomer focused on the $M1$ channel. However, it was shown in Ref.~\cite{PavloE2} that the $E2$ channel can have a significant and even dominant contribution for internal conversion and EB transitions for thorium ions. Here we confirm these results in the crystal environment and show that for the dominant EB processes, the nuclear $E2$ pathway accounts for upwards of $85$\% to the final transition rate.

The second purpose of this work is to discuss the prospect of experimental implementation for the defect-state-based EB processes and the resulting solid-state nuclear clock performance. The starting point here is the precise identification of the defect state energy and width, which could be performed in VUV fluorescence or absorption measurements. For defect energies approaching the band gap, the direct spectroscopic detection is mainly limited by the doped crystal transparency. In addition, it is compulsory to investigate possible broadening mechanisms of the defect states otherwise difficult to model theoretically. Finally, the nuclear clock performance based on quenching of the isomer population via driven EB channels is investigated theoretically. Our results show that the quenching can improve the short-term stability of the clock by more than one order of magnitude. 

The paper is structured as follows. In Sec.~\ref{sec:EB} the formalism of both spontaneous and driven EB processes in the crystal environment are presented in the non-relativistic limit. Details regarding state parity and allowed transitions are discussed for $E1$, $M1$ and $E2$ bridge processes. The density functional theory methods used for the calculation of electronic defect states  are presented in Sec.~\ref{sec:DFT}. Numerical results are presented in Sec.~\ref{numres}, including a discussion of convergence criteria for the EB calculations in Sec.~\ref{sec:conv}. The impact of the nuclear $M1$ and $E2$ channels are discussed in the context of EB processes showing their relative strength.  Section \ref{sec:exp} discusses experimental approaches for the precise  measurement of the electronic defect states in the crystal, along with potential difficulties. Section \ref{sec:clock} investigates the potential impact of driven EB schemes as means of isomer population quenching on the performance of a solid-state nuclear clock. Concluding remarks are given in the final Sec.~\ref{concl}.

\section{Electronic bridge in the crystal environment \label{sec:EB}}

\begin{figure}[htb!]
	\centering
	\includegraphics[scale=1]{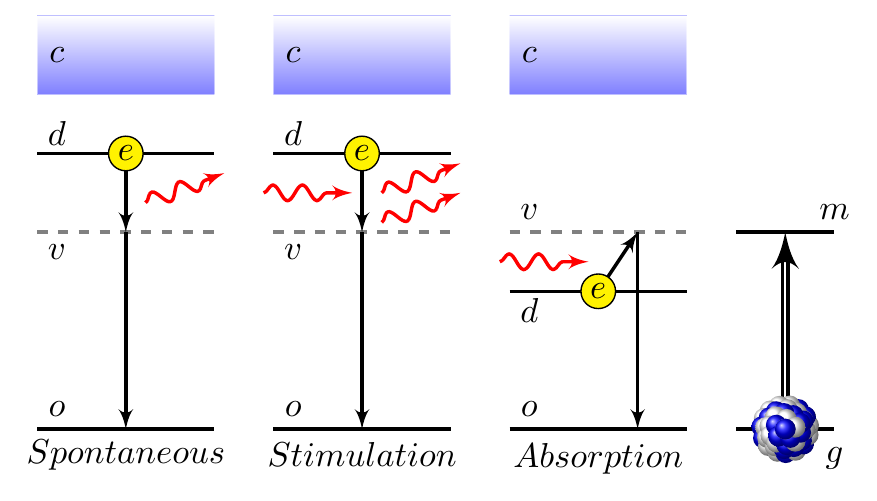}\\
	\caption{EB process for the excitation of $^{229m}$Th from the ground state $|g\rangle$ to the isomeric state $|m\rangle$ (right graph) \cite{Nickerson20PRL}. The initially populated electronic defect states $|d\rangle$ lie in the crystal band gap  above or below the isomer energy. The EB process occurs either spontaneously (left graph) or assisted by an optical laser in the stimulated or absorption schemes (middle graphs).  In all cases, EB proceeds via a virtual electronic state $|v\rangle$ and ends in the ground state $|o\rangle$, where the conduction band states are given by the set $|c\rangle$.}
	\label{fig:stimEB}
\end{figure}

The term EB is used in the literature for both nuclear excitation and nuclear decay facilitated by the coupling to the atomic shell. While electronic and nuclear transitions happen simultaneously, their energy does not have to match exactly; the difference in energy is carried away by or supplied by an emitted or absorbed photon, respectively. 
In the context of VUV-transparent crystals,  possible EB excitation schemes involving the excitation of the $^{229}$Th nucleus from the ground state $|g\rangle$ to the isomeric state $|m\rangle$ are illustrated in Fig.~\ref{fig:stimEB} \cite{Nickerson20PRL}. The VUV-transparent CaF$_2$ crystal presents a band gap of approx. 11.5 eV between the ground state $|o\rangle$ and the conduction band $|c\rangle$. Due to thorium doping,  electronic defect states $|d\rangle$   located in the range of the nuclear isomer appear in the crystal bandgap. The precision of DFT calculations is not sufficient to be confident whether the defect states are slightly above or slightly below the isomer. We therefore consider both possibilities in the following.

A spontaneous EB exciting the nuclear isomer can occur when the defect states  $|d\rangle$ are initially populated and  lie higher in energy than the isomeric state. This situation is illustrated in the left-most panel of Fig.~\ref{fig:stimEB}. The initially populated electronic defect states can decay to the ground state $|o\rangle$ by transferring the excitation energy to the nucleus. The process proceeds via a virtual electronic state $|v\rangle$ and the surplus of energy is emitted in the form of a photon. One can additionally stimulate the spontaneous process by shining a laser with the same frequency and polarization as the one of the outgoing photon. Should the defect states lie below the isomer, the spontaneous process is not possible. However, by providing the system with the missing energy in the form of a laser photon, absorption can render the EB energy transfer possible. In this case, the simultaneous decay of the defect state and absorption of the laser photon will lead to nuclear excitation and population of the isomer. 

The allowed transitions in the electronic shell, together with the nuclear transition multipolarity determine the multipolarity of the emitted photon.  In $^{229}$Th, the nuclear transition from the ground state $|g\rangle$ with angular momentum $5/2^+$ and positive parity to the isomeric state $3/2^+$ can proceed via  $M1$ and $E2$ multipole mixing. Thus, typically an allowed $E1$ transition between the initial and final electronic states will convert to an $E1$ multipolarity of the emitted photon.  When selection rules forbid the $E1$ channel for the EB photon, the much slower magnetic dipole or electric quadrupole channels should be considered. In the crystal environment, however, all electronic states are no true eigenstates of angular momentum or parity, and thus no  selection rules can be directly applied. In the following we present the application of the EB theoretical formalism to the crystal environment and discuss our knowledge of the defect states. 

\subsection{EB theoretical formalism}

As introduced in \cite{Nickerson20PRL} and presented in Fig.~\ref{fig:stimEB}, EB processes can be assisted by an optical laser which couples the initial or final electronic state with the virtual state causing stimulation or absorption and faster EB rates. For this we note once again here that the rate  $\Gamma^{st}(a\rightarrow b)$ of a laser-stimulated generic process $|a\rangle\rightarrow |b\rangle$ can be related to the rate of the corresponding spontaneous process $\Gamma^{sp}(a \rightarrow b)$ as \cite{LL_QED_1982, Sobelman_book_1979}
	\begin{align}
		\Gamma^{st}(a\rightarrow b) &= \Gamma^{sp}(a \rightarrow b) \frac{\pi^2c^2\hbar^2}{E^3} I\, ,
		\label{eqn:first}
	\end{align}
where the spectral intensity of the laser source $I$ is given  in SI units as W/(m$^2$s$^{-1}$). The required photon energy is denoted by $E=\hbar\omega_{ab} = \hbar (\omega_a - \omega_b)$, and $c$ stands for the speed of light. Via detailed balance, the stimulated rate $\Gamma^{st}(a\rightarrow b)$ can be related to the inverse absorption process rate as $ \Gamma^{ab}(b\rightarrow a) = \Gamma^{st}(a\rightarrow b) \delta(a\rightarrow b)$, with  $\delta(a\rightarrow b) = N_a/N_b$ the ratio of multiplicities of sets $\{|a\rangle\}$ versus $\{|b\rangle\}$. Hence, as an input we must first calculate the spontaneous EB process of interest. Referring to Fig.~\ref{fig:stimEB}, for the \emph{Absorption} case we can connect the spontaneous and laser-assisted processes by considering the time-reversed picture, i.e., by reversing the initial and final states of the electron and nucleus along with the direction of flow of the photon and transition arrows.  

For the expression of the spontaneous EB rates, we switch to atomic units ($\hbar=m_e=e=1$). Depending on the multipolarity of the emitted photon, we can write the expressions for $E1$, $M1$ and $E2$ bridge rates as, 
\begin{eqnarray}
\Gamma_{E1}^{sp}& =& \frac{4}{3}\left(\frac{\omega_p}{c}\right)^3 \frac{1}{N_g N_d}\sum_{\substack{m,g,\\o,d}}|\langle m,o|\widetilde{\boldsymbol Q}_{E1}|g,d\rangle|^2\,,
	 \label{eqn:EBsponexc1_cry} \\
\Gamma_{M1}^{sp} &=& \frac{4}{3}\frac{\omega_p^3}{c^5} \frac{1}{N_gN_d}\sum_{\substack{m,g,\\ o,d}}|\langle m,o|\widetilde{\boldsymbol Q}_{M1}|g,d\rangle |^2,
\label{eqn:EBsponexc_M1} \\
\Gamma_{E2}^{sp}  &=& \frac{1}{15}\left(\frac{\omega_p}{c}\right)^{5}\frac{1}{N_gN_d}\sum_{\substack{m,g,\\ o,d}}|\langle m, o|\widetilde{\boldsymbol Q}_{E2}|g,d\rangle |^2
\label{eqn:EBsponexc_E2}\, .
\end{eqnarray}
States are denoted for example by $|g,d\rangle=|g\rangle|d\rangle$ where $g$ represents the quantum numbers of the nuclear ground state and $d$ that of the defect state. The ground state  $|o\rangle$ is taken as the highest energy valence band state. The sums over $d$ and $o$ are performed over the spin degenerate sublevels of each respective state. The frequency of the emitted photon is denoted by $\omega_p = \omega_{do} - \omega_{mg}$, and the degeneracies of the nuclear ground and defect state are given by $N_g$ and $N_d$, respectively.  The bridge operators $\widetilde{\boldsymbol Q}_{\mu L}$ are spherical tensor operators of type $\mu$ (electric $E$ or magnetic $M$), multipolarity $L$ and have  $2L+1$ spherical components. The bridge operator matrix elements can be written as
\begin{multline}
\langle m,o|\boldsymbol{\widetilde{Q}}_{\mu L}|g, d\rangle = \sum_{\lambda K,q}(-1)^q\left[\sum_{n} \frac{\langle o|\boldsymbol{Q}_{\mu L}|n\rangle\langle n|\mathcal T_{\lambda K,q}|d\rangle}{\omega_{dn}-\omega_{mg}}\right. \\
\left.+\sum_{k} \frac{\langle o|\mathcal T_{\lambda K,q}|k\rangle\langle k|\boldsymbol{Q}_{\mu L}|d\rangle}{\omega_{ok}+\omega_{mg}}\right]\langle m|\mathcal M_{\lambda K,-q}|g\rangle. \label{eqn:EBsponexc2}
\end{multline}
Here, $\lambda K$ represent the multipolarities of the coupling operators $\mathcal{T}_{\lambda K,q}$ and nuclear transition operators $\mathcal M_{\lambda K,-q}$ where $q= (-K, -K+1,\ldots, K-1,K)$ are their spherical components  \cite{Akhiezer_QED,Varshalovich_QTAM}. The summations are performed over all unoccupied intermediate electronic states denoted by $|n\rangle$ and $|k\rangle$.  The spherical tensor operator $\boldsymbol Q_{\mu L}$ describes the emitted photon of multipolarity $\mu L$. Please note that depending on $\mu L$, these operators have different dimension, corresponding to the different multiplication factors  in Eqs.~(\ref{eqn:EBsponexc1_cry}-\ref{eqn:EBsponexc_E2}), and also a different number of spherical components. 
In the case of an $E1$ bridge,  $\boldsymbol{Q}_{E1} = -\boldsymbol r$, where $\boldsymbol r$ is the position relative to the thorium nucleus which is considered the origin. In a similar fashion we have $\boldsymbol{Q}_{M1}  =-\frac{1}{2} (\boldsymbol{l} + \boldsymbol{\sigma})$ and $\boldsymbol{Q}_{E2} = -\sqrt{\frac{4\pi}{5}}r^2\boldsymbol{Y}_{2}$ for $M1$ and $E2$ EB processes, respectively. Here, $\boldsymbol l$  is the orbital angular momentum of the electron, $\boldsymbol \sigma$ are the Pauli matrices  and we use the notation  $\boldsymbol{Y}_2$ ($Y_{2,q}$) for the spherical harmonics.

The nuclear isomeric transition in $^{229}$Th is a mixture of magnetic dipole and electric quadrupole which restricts the sum over $\lambda K$ to these two multipolarities. This is not to be confused with the multipolarity $\mu L$ of the bridge photon which is either emitted or absorbed. In the non-relativistic limit, the magnetic-dipole coupling operator reads \cite{Abragam}
\begin{align}
	\mathcal{T}_{M1,q}= \frac{1}{c}\left[\frac{l_q}{r^3}-\frac{\sigma_q}{2r^3} + 3\frac{r_q (\boldsymbol \sigma \cdot \boldsymbol r)}{2r^5} + \frac{4\pi}{3}\sigma_q \delta(\boldsymbol r)\right],
	\label{eqn:TM1}
\end{align}
where $\boldsymbol l$ ($l_q$) is the orbital angular momentum of the electron, $\boldsymbol \sigma$ ($\sigma_q$) are the Pauli matrices (in spherical basis) and $\delta(\boldsymbol r)$ denotes the Dirac delta function. The electric-quadrupole coupling operator is given by \cite{Varshalovich_QTAM}
\begin{align}
	\mathcal{T}_{E2,q} &= -\frac{1}{r^3} \sqrt{\frac{4\pi}{5}}Y_{2,q}(\theta, \phi)\, .
\end{align}
	
An important ingredient for calculating the electronic matrix elements of $\boldsymbol{Q}_{\mu L}$ and $\mathcal T_{\lambda K,q}$ are the crystal wave functions for the valence, defect and conduction band states. These are obtained from DFT calculations, together with the corresponding energies $\omega_{dn}$ and $\omega_{ok}$. Our DFT approach and its limitations are presented in Sec.~\ref{sec:DFT}. The sums over intermediate states require a good knowledge of a large number of states in the conduction band. Our results on the convergence of the EB rates will be discussed in Sec.~\ref{sec:conv}.
	
Returning to the EB rate expression in Eq.~\eqref{eqn:EBsponexc2}, the last term on the right-hand side $\langle m|\mathcal M_{\lambda K,-q}|g\rangle$ stands for the matrix elements of the nuclear transition operators. These are connected via the  the Wigner-Eckart theorem \cite{Edmonds_AM} to the reduced transition probabilities   $B_\downarrow$ for which we use theoretical values predicted in  Ref.~\cite{Minkov_Palffy_PRL_2017}.

\subsection{Defect states in Th:CaF$_2$ \label{sec:DFT}}

CaF$_2$  has an experimentally measured band gap in the region of $11$-$12$ eV \cite{Rubloff1972, Barth1990, Tsujibayashi2002}. DFT calculations using the Vienna Ab initio Simulation Package VASP at the Gamma-point \cite{PAW, PhysRevB.59.1758} show that upon doping with thorium, there are eight spin-degenerate defect states $\{|d \rangle \}= \{|d_1\rangle, \ldots, |d_8\rangle\}$ appearing within the band gap of undoped CaF$_2$. These states are localized on the Th dopant and its 5f orbital, while the transition from the valence band $|o\rangle$ to the set $\{|d\rangle\}$ is reminiscent of a $2p$ orbital electron of an interstitial fluorine ion migrating to the Th ion. For the DFT calculations we use the Heyd-Scuseria-Ernzerhof hybrid functional (HSE) \cite{HSE, hse06} which is an improvement to other generalized gradient approximations for the description of various physical properties, especially for the band gap~\cite{PhysRevB.84.075205}. Depending on the case under investigation, HSE is otherwise at least en-par in terms of performance and quality to other hybrid methods~\cite{Garza2016, Das2019}.

\begin{figure}[htb!]
	\centering
	\includegraphics[scale=1]{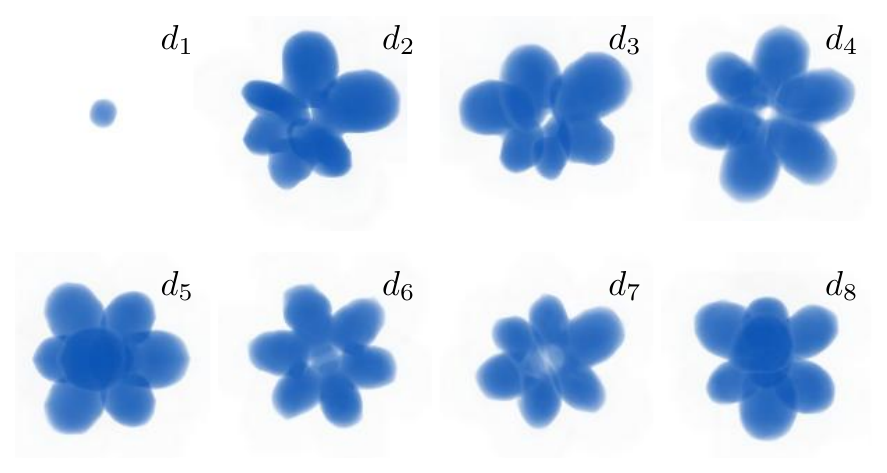}
	\caption{Electron density illustrations  for the defect states labeled  $\{|d\rangle\} = \{|d_1\rangle,\ldots, |d_8\rangle\}$.}
	\label{fig:wavefunctions}
\end{figure}

DFT provides one-electron wave functions and energies for the defect states and for the valence and conduction bands of the crystal. Figure~\ref{fig:wavefunctions} displays the electron density of the eight defect states localized around the thorium nucleus in the crystal unit cell. Our DFT calculations underestimated the band gap of undoped CaF$_2$ by approx.~17\% as compared to experimentally measured values. In order to match this calculated band gap  with the experimentally measured value of $11.5$ eV, a scaling procedure via the scissors operator is applied in the calculation \cite{Godby88, Levine89}. As a result, the (scaled) defect states lie in the region of $10.5$ eV. The obtained energy values are presented in Sec.~\ref{numres} in Table~\ref{table:decay}. We emphasize here that we cannot  undoubtedly assign the defect states' energy  without further experimental investigation. As such, energies given by DFT\&S(cissor)  should only be understood as an estimate, and will be used along with  energy scalings employed to better understand the EB choices in the energy region around 8~eV. 

Since VASP uses the Projector Augmented Wave (PAW) method~\cite{PhysRevB.50.17953}, the all-electron Kohn-Sham (AE-KS) wave function $|\Psi\rangle$ near the nucleus is augmented in order to increase numerical performance. This augmentation applies a linear operator $\mathcal{O}$ to the so-called pseudo wave function $|\tilde{\Psi}\rangle$ such that $|\Psi\rangle = \mathcal{O} |\tilde{\Psi}\rangle$. The linear operator $\mathcal{O}$ is defined as $\mathcal{O} = 1 + \sum_i \left( |\phi_i\rangle - |\tilde{\phi}_i\rangle \right) \langle\tilde{p}_i|$, where $|\phi_i\rangle$ and $|\tilde{\phi}_i\rangle$ are the AE- and pseudo partial waves respectively and $\langle \tilde p_i|$ are the projectors. 

In this work we compute the matrix elements in Eq.~\eqref{eqn:EBsponexc2} in the basis of one-electron states using a real space representation of $|\Psi\rangle$. This representation was obtained by extracting the projectors, partial waves and pseudo wave functions from VASP and carrying out the linear transformation $\mathcal{O}:|\tilde\Psi\rangle \rightarrow |\Psi\rangle$. We estimate the accuracy of the resulting AE-KS wave function $|\Psi\rangle$ by calculating its norm, which is related to the pseudo wave function via $\langle \Psi |\Psi \rangle = \langle \tilde{\Psi} |S| \tilde{\Psi} \rangle$. Here, $S = 1 + \sum_{ij} |\tilde{p_i}\rangle \left( \langle \phi_i|\phi_j\rangle- \langle \tilde{\phi}_i|\tilde{\phi}_j\rangle\right) \langle \tilde{p_j}|$. We find for the difference $\left| \langle \Psi |\Psi \rangle - \langle \tilde{\Psi} |S| \tilde{\Psi} \rangle \right| < 3\%$, suggesting that our procedure has only minor numerical errors. 

Due to the Hohenberg-Kohn theorem~\cite{PhysRev.136.B864}, DFT is only valid for the ground state. When an electron is excited into a defect state or beyond, the energy of those states is subject to change due to dynamic effects such as the electron-hole interaction. An estimation of the strength of this effect would require a calculation which includes these correlations, such as the GW-method~\cite{GW1} (Green's function $G$ and screened Coulomb interaction $W$), where the exchange correlation potential is replaced by the many-body self energy~\cite{GW2}, or other approaches of quantum chemistry. Such an investigation will be reserved for future efforts once more information is known experimentally about the thorium defect states in question.     

The last term in the $M1$ coupling operator in Eq.~\eqref{eqn:TM1}, requires the value of the electronic wave functions at the position of the $^{229}$Th nucleus. VASP uses a radial grid on exponentially spaced grid points excluding the atom center. To obtain the value of the wave function at the $^{229}$Th nucleus, the one-electron wave functions  are fit using the function $\Psi_{\text{fit}}(r) = \Psi_0 \exp(-rb)$, where $\Psi_0$ and $b$ are the fit parameters. This ansatz is well justified for non-relativistic s-like orbitals at small $r$ values. With increasing radial distance the wave function becomes less dominantly $s$-like. To account for this we define a maximum distance to the nucleus for further considerations. We choose this length to be half the distance of the first extreme value of the wave function in each radial direction, since only states with $l \neq 0$ can produce such points.  For each pair of spherical coordinates $\phi$ and $\theta$, we construct a fit with parameters $\Psi_0$ and $b$. All $\Psi_0$ parameters for these fits in radial direction must converge for the wave function to be well defined. The final value for $\Psi(r = 0)$ is then the mean of all $\Psi_0$ values.

\section{Numerical results \label{numres}}

In the following we present our numerical results for the EB rates, investigating both different bridge photon multipolarity channels, as well as the individual contributions of the nuclear $M1$ and $E2$ decays. For the DFT\&S calculation we have used  a unit cell of 66 fluorine, 31 calcium and a single thorium atom. Since the  wave functions of electrons in the crystal environment are not eigenstates of either angular momentum or parity, the spatial parts of the wave functions are   only defined by their energy.	 Wave functions were calculated on a spherical grid with the number of points $(N_r, N_\theta, N_\phi) = (353,29,60)$, considering constant spacing in angular components and the spacing in the radial component followed $r_n=r_0\,e^{n/\kappa}$ with $r_0=1.35\times10^{-4}\,a_0$ and $\kappa = 31.25$. Spherical grids as large as $(N_r, N_\theta, N_\phi) = (353,44,90)$  were tested but did not improve the accuracy of the result significantly. The calculated and scaled (via the scissor operator procedure) defect state energies $E_d$ are presented in the second column of Table \ref{table:decay}. 
\bgroup
\def\arraystretch{1.5}
\setlength\tabcolsep{2mm}
\begin{table}[htb!]
\begin{tabular}{ccccc}
\hline\hline
& \multirow{2}{4em}{$E_d$ [eV]} & \multicolumn{3}{c}{$A^{sp}(d \rightarrow o)$ [s$^{-1}]$}  \\ \cline{3-5}
 &	&$E1$  &$M1$  &$E2$\\  \hline \\[-9pt]
$|d_1\rangle$ 	& 9.90 &	 $7.84\times 10^{4}$ & $5.04\times 10^{2}$ & $9.26\times 10^{1}$   \\
$|d_2\rangle$	& 10.43 & 	$4.35\times 10^{0}$ & $2.51\times 10^{1}$ & $2.57\times 10^{0}$   \\ 
$|d_3\rangle$	& 10.50 &	$1.99\times 10^{6}$ & $7.09\times 10^{1}$ & $5.95\times 10^{1}$   \\
$|d_4\rangle$	& 10.51 & 	$6.16\times 10^{1}$ & $4.81\times 10^{1}$ & $5.69\times 10^{0}$   \\ 
$|d_5\rangle$	& 10.59 &	$7.27\times 10^{6}$ & $6.64\times 10^{1}$ & $3.03\times 10^{1}$   \\ 
$|d_6\rangle$	& 10.63 & 	$1.12\times 10^{5}$ & $2.20\times 10^{1}$ & $1.82\times 10^{0}$   \\ 
$|d_7\rangle$	& 10.68 & 	$1.19\times 10^{7}$ & $1.24\times 10^{1}$ & $1.56\times 10^{1}$   \\ 
$|d_8\rangle$	& 11.01 & 	$2.16\times 10^{5}$ & $2.18\times 10^{1}$ & $2.74\times 10^{2}$   \\ \hline \hline
\end{tabular}
\caption{HSE defect state energies $E_d$ obtained from DFT\&S and electronic transition rates $A^{sp}(d \rightarrow o)$  from the defect state to the ground state calculated using the $E1$, $M1$ and $E2$ multipole operators, respectively.  }
\label{table:decay}
\end{table}
\egroup

All EB schemes under investigation (see Fig.~\ref{fig:stimEB}) consider as initial state one of the defect states. The latter can be reached by VUV excitation. It is therefore useful to start by calculating the matrix elements $\langle o|\boldsymbol{Q}_{\mu L}|d\rangle$ for  $\mu L=E1$, $M1$ and $E2$.  The  corresponding electronic decay rates $A^{sp}_{\mu L}(d\rightarrow o)$ are presented in Table \ref{table:decay}, and calculated by the corresponding equations \eqref{eqn:EBsponexc1_cry}, \eqref{eqn:EBsponexc_M1}, \eqref{eqn:EBsponexc_E2}, where $|\langle m,o|\widetilde{\boldsymbol Q}|g,d\rangle| \rightarrow |\langle o|{\boldsymbol Q}|d\rangle|$. It is this rate which also determines which of the defect states is most likely to be excited by our initial excitation, and the favoured multipolarity.  For most of the defect states, the $E1$ decay is dominant, and $|d_7\rangle$ has the largest decay rate. Correspondingly, we expect that $|d_7\rangle$ is the easiest level to excite from the ground state $|o\rangle$ via VUV laser pumping. In the case of $|d_2\rangle$ and $|d_4\rangle$ the $M1$ and $E2$ contributions are the same order or larger than the $E1$ one; however, these states should be seldomly populated by the initial excitation in favour of the faster rates of other states such as $|d_7\rangle$. 

In order to estimate the population of the initial electronic state, i.e., of the defect states, we obtain the steady state solution of the Bloch equation 
\begin{equation}
\dot{\rho}_d = \rho_o A^{ab}_{E1}(o\rightarrow d) - \rho_d \left[A^{sp}_{E1}(d\rightarrow o) + A^{st}_{E1}(d\rightarrow o)\right]
\label{Bloch}
\end{equation}
where $A^{ab}_{E1}(o\rightarrow d)$ and $A^{st}_{E1}(d\rightarrow o)$ are the absorption and stimulated decay rates for the transition  $o\rightarrow d$  in the presence of a VUV laser field with intensity $I_d$, following the recipe of Eq.~\eqref{eqn:first}. The equation above is used in the following to derive the population $\rho_{d_i}$ of individual defect states $|d_i\rangle$. In addition, for a crude approximation, we calculate also average EB rates which consider the complete set of defect states $\{|d\rangle\}$ as quasi-degenerate levels. In this case, the rates $ A^{ab}_{E1}(o\rightarrow \{d\})$,    $ A^{sp}_{E1}( \{d\}\rightarrow o)$ and $ A^{st}_{E1}( \{d\}\rightarrow o)$ in Eq.~\eqref{Bloch} are calculated according to Eqs.~\eqref{eqn:EBsponexc1_cry}, \eqref{eqn:EBsponexc_M1}, \eqref{eqn:EBsponexc_E2} with the substitution $|\langle m,o|\widetilde{\boldsymbol Q}|g,\{d\}\rangle| \rightarrow |\langle o|{\boldsymbol Q}|\{d\}\rangle|$, allowing the sum over $d$ to run over all defect states $d_i\in\{|d\rangle\}$ and further considering the photon energy $\omega_p$ factor, in this case $d_i$-dependent, under this summation. As a result, Eq.~\eqref{Bloch} delivers in this case an average defect state population, which we then use to obtain approximate average EB rates. 

The total EB rate achieved in the crystal is given by multiplication with the population of the initial state $\rho_d$, and the number of nuclei in the crystal exposed to the excitation process $N$, giving (once more in SI units)
\begin{align}
	N\rho_d\Gamma^{st}(|g,d\rangle \rightarrow &|m,o\rangle) \approx \nonumber\\
	&\frac{NN_d(\pi c\hbar)^4}{N_oE_d^3 E_p^3} I_dI_p\Gamma^{sp}(|g,d\rangle \rightarrow |m,o\rangle)\, ,\\
	N\rho_d\Gamma^{ab}(|g,d\rangle \rightarrow &|m,o\rangle) \approx \nonumber\\
	&\frac{NN_m(\pi c\hbar)^4}{N_gE_d^3 E_p^3} I_dI_p\Gamma^{sp}(|m,o\rangle \rightarrow |g,d\rangle)\, ,
\end{align}
where for simplicity we have assumed $I_d\ll \frac{E_d^3}{\pi^2 c^2 \hbar^2}$ and $\rho_o=1$ for the start of the excitation process. 

The two laser intensities appear as multiplication factors in the two equations above. We recall that $I_d$ refers to the source used to excite the electronic shell to the defect state $|o\rangle \rightarrow |d\rangle$, while $I_p$ is the intensity of the optical source used to drive the desired electronic bridge process $|g,d\rangle \rightarrow |m,o\rangle$ by coupling with the virtual state $|d\rangle \rightarrow |v\rangle$. The notation $E_p$ is used for the photon energy of the optical laser driving the bridge scheme. Furthermore, $N_o$ and $N_m$ are the degeneracies of the electronic ground and nuclear isomeric states, respectively.

\subsection{Convergence \label{sec:conv}}

As seen in Eq.~\eqref{eqn:EBsponexc2} the final rate requires a summation over all unoccupied intermediate states.  The conduction band $\{|c\rangle\}$ offers an infinite set of possible intermediate states, and the denominators  in Eq.~\eqref{eqn:EBsponexc2} are only slowly suppressing their contributions. Increasing the number of intermediate states should therefore be continued until convergence is reached. As an example, we will consider such convergence using the system energies given by DFT\&S. All the defect state energies (see Table~\ref{table:decay}) lie in this case above the isomer energy. 

\begin{figure}[!htb]
	\centering
	\includegraphics[scale=1]{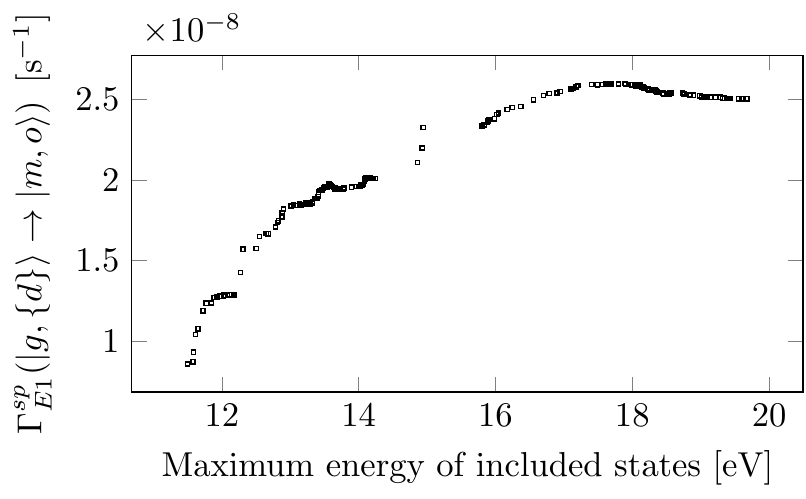}
	\caption{Average spontaneous EB rate $\Gamma^{sp}_{E1}$ as a function of the maximum energy of the included states measured with respect to the electronic ground state $|o\rangle$. }
	\label{fig:conv1}
\end{figure}

We start by calculating the spontaneous $E1$ bridge rate $\Gamma^{sp}_{E1}(|g,\{d\}\rangle\rightarrow |m,o\rangle)$, where the initial electronic state is taken as the set of eight spin-degenerate defect states. The nuclear matrix element in Eq.~\eqref{eqn:EBsponexc2} is calculated using the theoretically predicted values (in Weisskopf units, W.u.) $B_W(M1, m\rightarrow g) = 0.0076 \,\,\,\text{W.u.}$, $B_W(E2, m\rightarrow g) = 27.04 \,\,\,\text{W.u.}$ \cite{Minkov_Palffy_PRL_2017} for the reduced transition probabilities. 
The rate $\Gamma^{sp}_{E1}(|g,\{d\}\rangle\rightarrow |m,o\rangle)$ is plotted in Fig.~\ref{fig:conv1} as a function of the maximum energy of the included states with respect to the highest energy valence band, 	i.e., the electronic ground state $|o\rangle$. Convergence is achieved with $\Gamma^{sp}_{E1}(|g,\{d\}\rangle\rightarrow |m,o\rangle)\approx 2.5\times 10^{-8}$ s$^{-1}$ and the order of magnitude of the rate is stable throughout the entire range. With increasing energy the conduction band states become less accurate as electron-hole interactions are neglected. However, due to the convergence within the order of magnitude, we expect this error to be to be inconsequential for our purposes.  

\begin{figure}[!htb]
	\centering
	\includegraphics[scale=1]{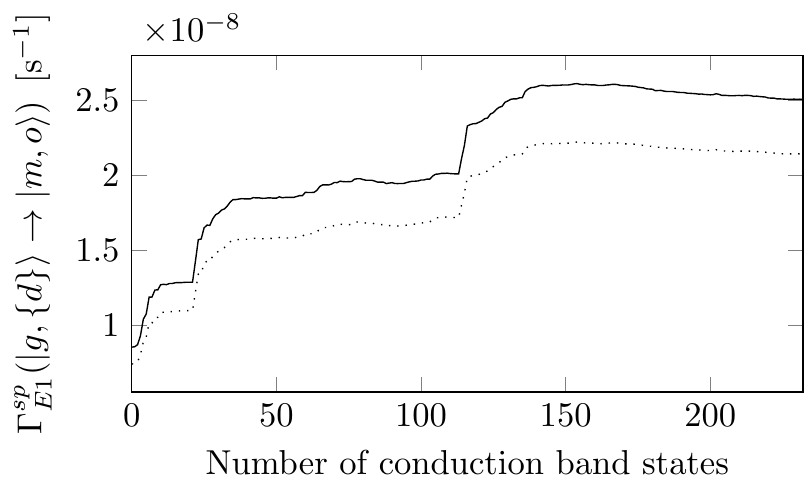}
	\caption{Average spontaneous EB rate $\Gamma^{sp}_{E1}$ as a function of number of conduction band states included in the intermediate summation of $|n\rangle$ and $|k\rangle$ seen in \eqref{eqn:EBsponexc2}. Solid line is the total rate, while the dotted line shows the $\boldsymbol{\mathcal{T}}_{E2}$ contribution alone. }
	\label{fig:conv}
\end{figure}

%
Additionally, $\Gamma^{sp}_{E1}(|g,\{d\}\rangle\rightarrow |m,o\rangle)$ is plotted in Fig.~\ref{fig:conv} as a function of number of conduction states included in the sum over the intermediate states $|n\rangle$ and $|k\rangle$ in Eq.~\eqref{eqn:EBsponexc2}. We use lines instead of points in the graph to more clearly illustrate  the contribution of the $E2$ nuclear decay channel as discussed in Sec.~\ref{sec:coupling}. 
Note that each conduction band state $|c\rangle$ is spin degenerate such that the total number of states accounting for degeneracy is twice as much as that shown on the $x$-axis of Fig.~\ref{fig:conv}. The maximum number of spin-degenerate states considered in the set $\{|c\rangle\}$ is 232.  

Few comments are appropriate at this point. By examining equation \eqref{eqn:EBsponexc2} it is clear that as the energy difference between the intermediate states and the initial and final electronic states increases, the contribution to the rate decreases. The sum of the denominator alone is simply the harmonic series which cannot result in convergence.  As such, the numerator must also plummet to zero. When considering transitions in a single atom, it is expected that with increasing energy difference the wave function overlap will typically decrease, resulting in an ever smaller numerator. However, the different shapes of atomic orbitals would prevent a completely smooth convergence of the summation. This is even more so in the crystal environment.  Although the general trend of decreasing wave function overlap with increasing transition energy holds, this is not necessarily smooth. At particular energies, electronic transitions  between states with more localized wave functions on neighbouring ions may occur. Such transitions can have larger overlap and bring (large) positive or negative contributions, resulting in a visible upwards or sometimes downwards step in the total rate.  

This step-like behaviour can be observed at several conduction band state energies, in particular around $11.7$ eV, $12.3$ eV, or $15$ eV. The steps become even more obvious in gaps in between conduction band energies (as calculated for the Gamma-point). This is the case, for example, for the three conspicuous data points around 15 eV in Fig.~\ref{fig:conv1} resulting in an upwards step in the rate, also seen at conduction band number 115 in Fig.~\ref{fig:conv}. The three points correspond to three conduction band states which are particularly localized around the impurity consisting of the Th and interstitial F ions. Transitions between these and the set $\{|d\rangle\}$ result in large contributions via matrix elements of the operators $\boldsymbol{\mathcal{T}}_{\lambda K}$ and $\boldsymbol{Q}_{\mu L}$, i.e., large numerators in the respective summation terms and therefore a visible increase of the EB rate. Before concluding this part we should point out once more the limitations in our calculation, which is not independent of the chosen crystal cell size. Once states in the conduction band region are populated, electron-hole interactions not included in the calculation might qualitatively change the interpretation presented above.

\subsection{Dominant nuclear $E2$ channel \label{sec:coupling}}

For the calculation in Fig.~\ref{fig:conv} we have considered separately the two possible multipolarities for the nuclear transition, $M1$ and $E2$. For radiative decay of the isomeric state, the $M1$ component dominates by many orders of magnitude. However, for transitions mediated by the electronic shell, cases have been found where the $E2$ component is not negligible \cite{PavloE2}. For the present calculation, the nuclear $E2$ component turns out to be dominant. The contributions to $\Gamma^{sp}_{E1}$ due to $\boldsymbol{\mathcal{M}}_{E2}$ (and $\boldsymbol{\mathcal{T}}_{E2}$) are shown in Fig.~\ref{fig:conv} as a dotted line. Throughout the entire range used to test convergence, the nuclear transition multipolarity $\lambda K = E2$  component made up $\approx 85\%$ of the total rate. The difference of approx.~$15\%$ is made up for by $\lambda K = M1$. 

\begin{figure}[!htb]
	\centering
	\includegraphics[scale=1]{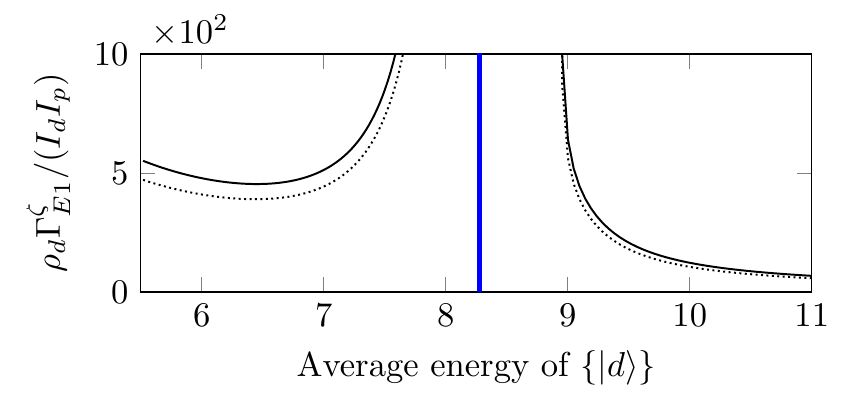}
	\caption{Normalized average driven EB rate $\rho_d\Gamma^{\zeta}_{E1}(|g,\{d\}\rangle \rightarrow |m,o\rangle)/(I_dI_p)$ [in units of m$^4$/(W$^2$s$^3$)] with $\zeta = ab/st$  as a function of average defect state energy. Solid line is the total EB rate, whereas the dotted line is the contribution from $\boldsymbol{\mathcal{M}}_{E2}$. The blue vertical line shows the isomer energy considered here, $E_m=8.28$~eV. Left (right) of this line, $\zeta=ab$ ($\zeta=st$). }
	\label{fig:E1avg}
\end{figure}

We now proceed to investigate the nuclear multipole contributions for the two laser-assisted schemes discussed in Fig.~\ref{fig:stimEB}. To this end we no longer use the fixed DFT\&S defect state energies given in Table~\ref{table:decay}, but allow the average energy of the set $\{|d\rangle\}$ to vary in the range $5-11$ eV by subtracting the same constant from each state energy. Also here we consider the initial electronic state as the average over the set of eight defect states. We calculate the $E1$ EB rate $\Gamma^{\zeta}_{E1}(|g,\{d\}\rangle \rightarrow |m,o\rangle)$, with $\zeta = st$ ($\zeta = ab$) for the range of average defect energy above (below) $E_m$. Figure~\ref{fig:E1avg} shows the total driven EB rates normalized to the intensities of the two lasers $I_d$ and $I_p$ as a function of average defect state energy along with the separate nuclear $E2$ coupling contribution to the rate. Once again, throughout the entire resonance energy range the nuclear $E2$ coupling component $\boldsymbol{\mathcal{M}}_{E2}$ is dominant with $\gtrapprox 85\%$. As such we confirm that the nuclear quadrupole channel is dominant when considering EB processes in $^{229}$Th:CaF$_2$ crystals. Further understanding of the nuclear processes in the crystal environment is expected once experiments confirm the energy and nature of the defect states.

\subsection{Comparison of bridge multipolarities}

\begin{figure}[!htb]
	\centering
	\includegraphics[scale=1]{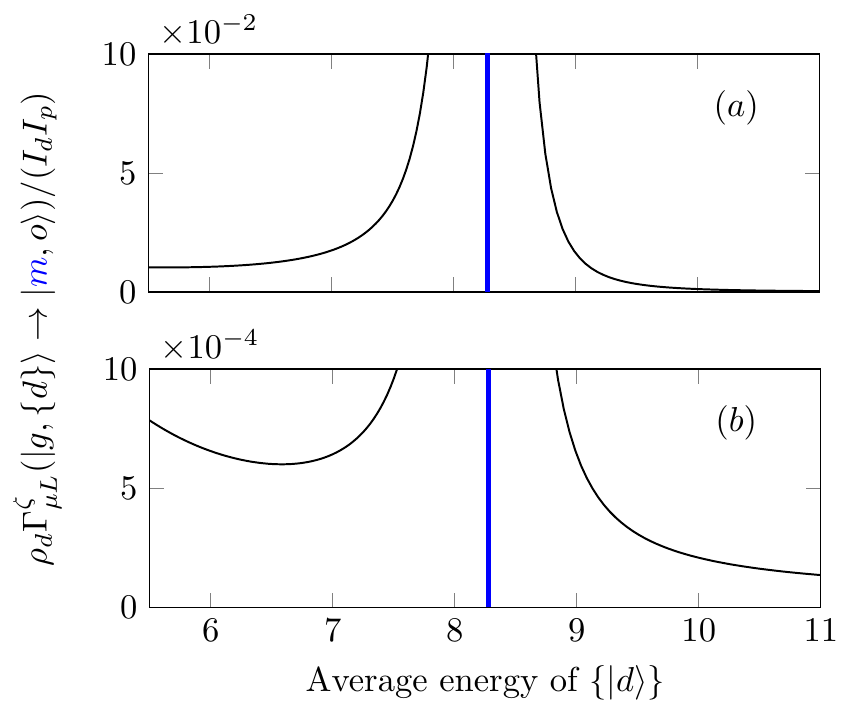}
	\caption{Normalized average driven EB rates $\rho_d\Gamma^{\zeta}_{\mu L}(|g,\{d\}\rangle \rightarrow |m,o\rangle)/(I_dI_p)$ [in units of m$^4$/(W$^2$s$^3$)], with $\zeta = ab/st$, as a function of average defect state energy where $(a)$ $\mu L= M1$ and $(b)$ $\mu L= E2$. The blue vertical lines mark the isomer energy $E_m=8.28$~eV. }
	\label{fig:M1E2avg}
\end{figure}

So far we have only considered bridge rates where the emitted or absorbed photon multipolarity was $E1$, given by $\Gamma^{\zeta}_{E1}$ with $\zeta=sp, ab, st$. Let us now focus on the $M1$ and $E2$ bridge multipolarities which can be calculated starting from Eqs.~(\ref{eqn:EBsponexc_M1}) and (\ref{eqn:EBsponexc_E2}). Figure~\ref{fig:M1E2avg} shows the  rates  $\Gamma^{\zeta}_{M1}$ and $\Gamma^{\zeta}_{E2}$ for the laser-diven EB processes for an initial averaged population of the defect states $\{|d\rangle\}$.
These rates can be directly compared to $\Gamma^{\zeta}_{E1}$ in Fig.~\ref{fig:E1avg}. As expected a similar resonant structure is seen, however with rates that are easily negligible in comparison to those seen with $\Gamma^{\zeta}_{E1}$.

More precisely, we can consider the rate resulting from a specific initial defect state. Referring to Table \ref{table:decay}, $|d_3\rangle$, $|d_5\rangle$ and $|d_7\rangle$ are the most easily populated via $E1$ excitation. Thus in Fig.~\ref{fig:d_3,5,7} we plot the EB rate $\Gamma^{\zeta}_{E1}(|g,d_i\rangle \rightarrow |m,o\rangle)$ where each of these states is individually taken as the initial electronic state.  Each displayed resonance corresponds to alignment in energy of one of the eight spin-degenerate defect states with the nuclear isomer.

\begin{figure}[!htb]
	\centering
	\includegraphics[scale=1]{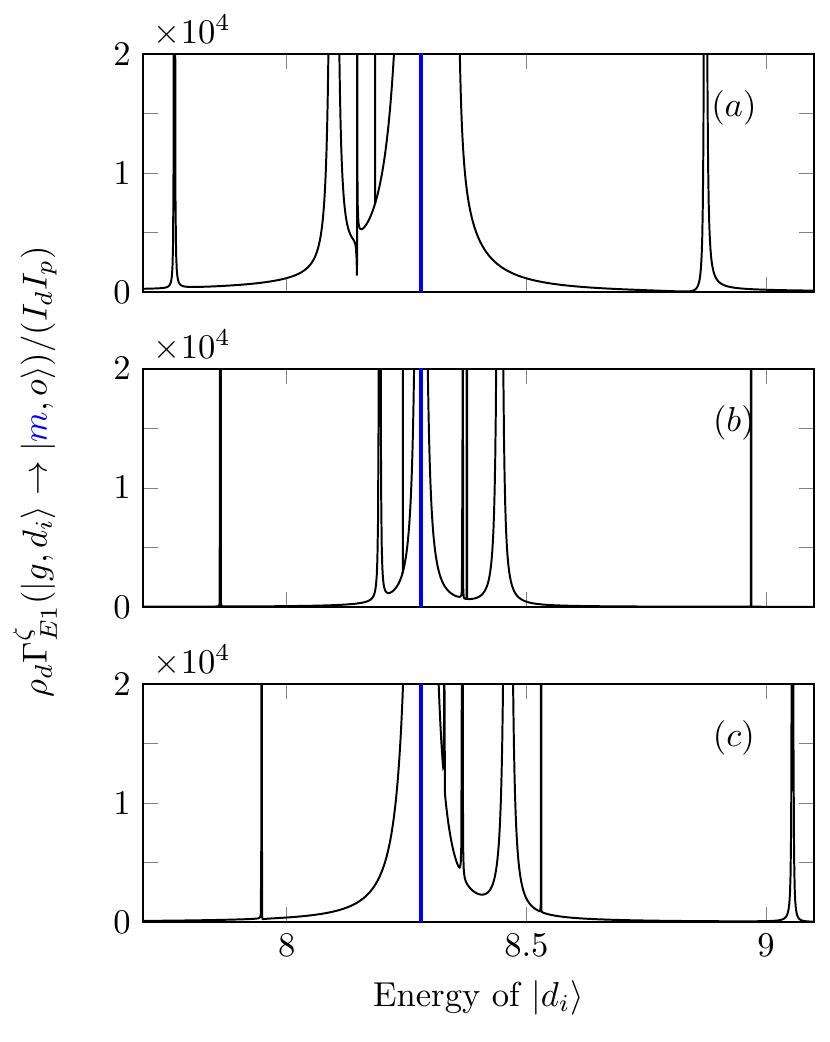}
	\caption{Normalized $E1$ bridge rates $\rho_d\Gamma^\zeta_{E1}(|g,d_i\rangle \rightarrow |m,o\rangle)/(I_dI_p)$ [in units of m$^4$/(W$^2$s$^3$)] as a function of initial defect state  $|d_i\rangle$ energy considering $(a)$ $i=3$, $(b)$ $i=5$, $(c)$ $i=7$. The blue vertical lines mark the isomer energy $E_m=8.28$~eV, with $\zeta=ab$ ($\zeta=st$)  left (right) thereof.}
	\label{fig:d_3,5,7}
\end{figure}

\begin{figure}[!htb]
	\centering
	\includegraphics[scale=1]{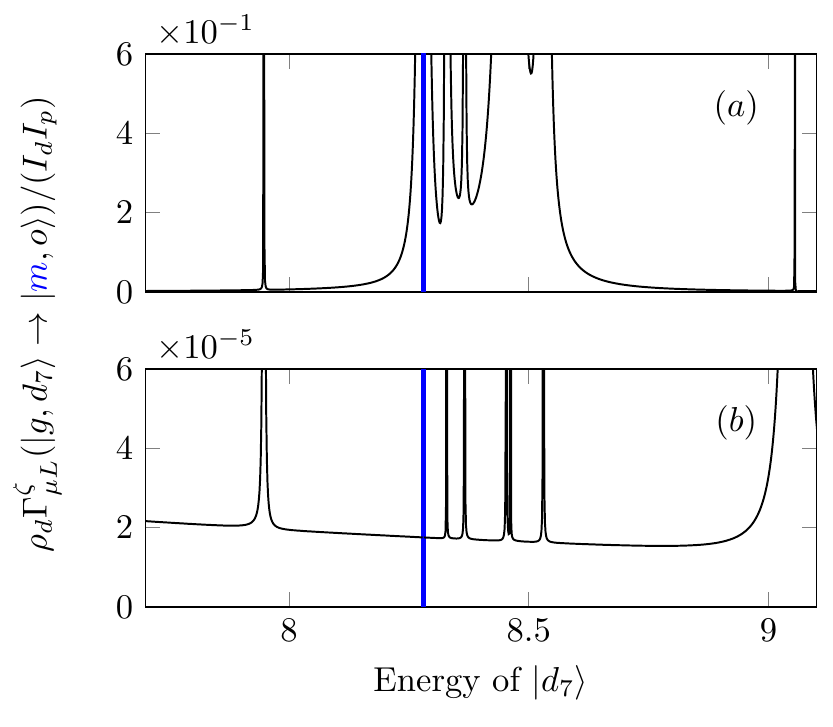}
	\caption{Normalized $(a)$ $\mu L=M1$, $(b)$ $\mu L= E2$ bridge rates $\rho_d\Gamma^\zeta_{\mu L}(|g,d_7\rangle \rightarrow |m,o\rangle)/(I_dI_p)$ [in units of m$^4$/(W$^2$s$^3$)] as a function of initial defect state energy. The blue vertical lines mark the isomer energy $E_m=8.28$~eV, with $\zeta=ab$ ($\zeta=st$)  left (right) thereof. }
	\label{fig:M1E2367}
\end{figure}

Figure~\ref{fig:M1E2367} shows the $M1$ and $E2$ bridge rates for the highest occupied state $|d_7\rangle$. Beyond the overall reduction in magnitude of the rates, we can also see how the relative widths of the individual resonances are affected by the change in allowed electronic transitions. Considering the orders of magnitude difference between the EB rates of different bridge photon multipolarity, we conclude that the $M1$ and $E2$ bridge rates can be safely neglected in this work. 

\section{Prospects of Experimental Implementation \label{sec:exp}}

The crystal environment offers a unique opportunity to investigate thorium at high densities. This system does, however, come with its own set of challenges including sources of background, laser damage, and the crystal's exciton spectrum.

Sources of background can be broadly categorized under the labels of photoluminescence and radioluminescence. Several of these sources have been studied specifically in Th:CaF$_2$ in Refs.~\cite{Stellmer2015,Stellmer2018, Dessovic_2014,Zimmermann_thesis_2010}.  Photoluminescence occurs from the excitation of unintended pathways in the crystal environment. These spurious excitations are the result of a variety of impurities introduced during the growth of the crystals as well as surface impurities introduced during storage and handling \cite{Denks2000, Ko2001}. Intrinsic to $^{229}$Th is the radioactive component of the background. $^{229}$Th undergoes $\alpha$-decay where the $\alpha$ particle and the $^{225}$Ra daugther nucleus have kinetic energies of $5.1$ MeV and $90$ keV, respectively. This energy release is then seen in the form of emitted photons as the particles crash through the crystal lattice up to distances of 30,000 and 30 lattice constants, respectively \cite{Stellmer2015}. Beyond the emitted photons forming a significant background to deal with, the damage caused by the high energy particles changes the structure of the crystal itself. Defects are left in the wake of the decay paths which change the available electronic transitions in their vicinity. The $\beta$-decay of thorium daughter products also contribute by causing Cherenkov radiation below 200 nm \cite{Stellmer2016}. 

Laser damage is an unavoidable consequence of dealing with VUV wavelengths. The exposure alone can cause damage to the crystal, which in some cases is reversible via annealing and tempering methods but can also cause an irreversible decrease in transparency by altering the crystal structure \cite{crystaldamage}. The impending laser damage usually occurs for exposure times ranging from minutes to months, however this is heavily dependent on the wavelength and intensity of irradiation, and also the prevalence and type of existing impurities in the crystal \cite{crystaldamage, Desovichthesis}.
 
The above mentioned sources of background and laser damage and impurities aside, one still has to work with the ideal transmission region of the CaF$_2$ crystal. This most important  transmission region results from a combination of the traditional band gap and the exciton absorption spectrum \cite{Tsujibayashi2002, Denk1999,  Denks2000, Bourdillont1976}. Pure CaF$_2$ exhibits absorption leading to exciton formation in the region above $10$ eV ($<125$ nm) which causes a sharp drop in transmission before the band gap energy is reached  \cite{Rubloff1972, Gorling2003, CF2data}. This impacts the experimental search for the Th-doping-induced defect states $\{|d\rangle\}$. Currently the DFT\&S scaling places the thorium defect states outside this transparency region, thus competing with the exciton spectrum which would pose issues for detection. To this end, the assignment of the thorium defect states in energy has yet to be studied experimentally. If the energies of the defect states lie in the transparency region of the Th:CaF$_2$  then various VUV spectroscopic methods can be used to detect and characterize them. Due to the small electronic transition rates $A^{sp}$, we propose VUV fluorescence spectroscopy for this search. The defect states will be excited either directly using VUV light \cite{Makhov_2008} or indirectly using x-ray \cite{Hidehiko_2011} or $\alpha$ excitation \cite{Furuya_2010}, as explored for the similar doped crystal Nd:CaF$_2$. After excitation, the time- and spectrally- resolved emission spectrum can be recorded. Another interesting factor is the potential broadening mechanisms of the electronic states, including those of the defect states under consideration here, which could lead to faster EB excitation and decay rates than the ones calculated based on DFT predictions. If a broadening mechanism beyond the DFT model causes the defect states to decay faster than fluorescence can be recorded, then VUV absorption spectroscopy could be used \cite{Hidehiko_2011}. The scaling procedure applied in Figs.~\ref{fig:E1avg}-\ref{fig:M1E2367}  will no longer be necessary once the energy of these states is known. At that point, the resonant structure of the system can be studied more precisely both experimentally and theoretically.

\subsection{EB quenching scheme and nuclear clock performance \label{sec:clock}}

\begin{figure}[htb!]
\includegraphics[scale=1]{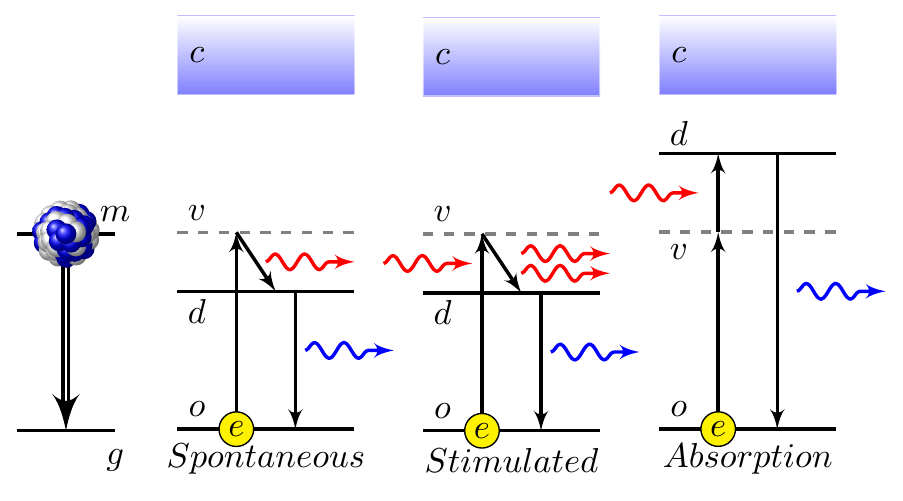}
\caption{Quenching processes for the de-excitation of $^{229m}$Th $|m\rangle \rightarrow|g\rangle$. The electronic defect states $|d\rangle$ can be thereby excited via a spontaneous (or additionally stimulated) EB scheme  provided  they lie below the isomer energy. If the defect states lie higher in energy than the isomer,  absorption of a laser photon is required. Wiggly arrows in red depict the EB photons related to the quenching of the isomeric state, either spontaneously emitted or externally pumped by a laser for the stimulated or absorption schemes. 
Photons in blue result from the subsequent spontaneous decay of the defect state  $|d\rangle \rightarrow |o\rangle$.}
\label{fig:quench}
\end{figure}

Once the thorium defect states are characterized, the focus then shifts to implementation of the available EB schemes and their impact on potential nuclear clock performance. For nuclear excitation, we have shown in Ref.~\cite{Nickerson20PRL} that using a VUV lamp  \cite{Stellmer2018} with $N \approx 3$ photons/(s$\cdot$Hz), a focus of $f=0.5$ mm$^2$ which corresponds to $I ={N\hbar\omega_{do}}/(2\pi f) \approx 1.6 \times 10^{-12}$ W/(m$^2$ s$^{-1}$) and a FWHM linewidth of $\approx 0.5$ eV, the EB  rate is more than 2 orders of magnitude faster than  direct photoexcitation. We now turn to inverse EB processes, spontaneous or optical-laser stimulated, which can be used to  quench the previously excited isomeric population \cite{Nickerson20PRL}. These processes are illustrated in Fig.~\ref{fig:quench}. With the nucleus initially in the isomeric state, a defect state situated lower in energy than $E_m$ can then be used for a spontaneous EB scheme that depletes the isomer. This happens via excitation of the electronic states and population of $|d\rangle$, where the energy mismatch is carried away by an emitted photon. In turn, the EB process can be stimulated by shining a laser with the frequency of this emitted photon. Should the defect states lie higher in energy than the isomer, a scheme using  absorption of an optical laser photon can be envisaged, as illustrated in the left-most panel of Fig.~\ref{fig:quench}. After the excitation of the defect states, these may decay radiatively, as depicted in Fig.~\ref{fig:quench} by the blue wiggled arrow. These isomer decay schemes can have much higher rates than the spontaneous radiative decay of the isomeric state, and may be used as ``managed quenching'' for preparation of clock states in a solid-state nuclear clock \cite{Kazakov_2012}. Instead of emission of the isomer transition photon, this laser-assisted quenching is accompanied by the photon from fast subsequent decay of the defect state, which may be used for detection of nuclear de-excitation.

As an example, we can consider the fixed energy case given by the DFT\&S defect state energies. The rate of the laser-assisted absorption quenching for this case was previously estimated as approx.~$\Gamma^{ab}_{qu}=0.07~{\rm s^{-1}}$ in Ref.~\cite{Nickerson20PRL},  which is 3 orders of magnitude larger than the spontaneous radiative decay rate $\Gamma\approx 10^{-4}~{\rm s^{-1}}$ \cite{Minkov_Palffy_PRL_2017}. This value was obtained using an optical laser intensity of $I_p=1$ W/(m$^2$s$^{-1}$). The quenching rate also follows equation~\eqref{eqn:first} and is thus  linearly dependent on the intensity of the driving laser. The largest variation in the quenching rate is likely to come from the experimental determination of the defect state energies which could place the quenching scheme closer to a resonance as discussed in earlier sections. 

Continuing with this example, let us estimate how the use of the laser-assisted EB quenching might improve the short-term stability of the solid-state optical clocks. Note that neither Rabi nor Ramsey interrogation schemes are applicable to such a clock, because of a huge difference between the short coherence time between the ground and the isomeric states (milliseconds) due to crystal lattice effects \cite{Kazakov_2012}, and much longer time necessary to bring the nuclei back into the ground state (tens of seconds even with laser-assisted EB de-excitation). Therefore, we consider the scheme based on counting the spontaneous (or laser-assisted) nuclear decay fluorescence photons after illuminating of the quantum discriminator with the VUV narrow-band laser. The excitation scheme as well as interrogation protocol considered below follows Ref.~\cite{Kazakov_2012}, up to replacement of the counting of nuclear fluorescence photons by counting of the photons from the decay of the defect state $|d\rangle \rightarrow |o\rangle$.

Consider first the excitation of the isomer transition in the crystal lattice environment under the action of a narrow-band VUV laser. This step is paramount for any nuclear clock, whether using trapped Th ions or Th-doped crystals. In the crystal  environment, 
the $^{229}$Th nuclei are subject to electric field gradients causing quadrupole splitting of the order of few hundred MHz \cite{Dessovic_2014, Pimon20}. In the absence of any external magnetic field, the quadrupole structure is degenerate with respect to the sign of projection of the nuclear angular momentum. To this end we consider stabilization of the laser on the pair of transitions between the states $|g_{1,2}\rangle=|{\rm ^{229g}Th,}\,I=5/2,\pm 3/2 \rangle$ and $|m_{1,2}\rangle=|{\rm ^{229m}Th,}\,I=3/2,m=\pm 1/2 \rangle$. Averaging over possible spatial orientations of the electric field gradient, one may obtain the equation for the total population $\rho_{\rm exc}$ of both the excited states $|m_{1}\rangle$ and $|m_{2}\rangle$ as 
\begin{equation}
\dot{\rho}_{\rm exc}=\frac{R}{1+\Delta^2/\gamma^2}
  -\left[\Gamma n^3 + \frac{5}{2}  \frac{R}{1+\Delta^2/\gamma^2} \right]
  \rho_{\rm exc},
\label{eq:1}
\end{equation}
where $\gamma$ is the relaxation rate of the nuclear transition coherences, primarily determined by the interaction with fluctuating fields inside of crystal, particularly, random magnetic field generated by the fluorine spins surrounding the thorium nucleus \cite{Kazakov_2012}.  The spontaneous radiative decay rate of the isomer is $\Gamma=10^{-4}~{\rm s^{-1}}$ (calculated using $B_W(M1, m\rightarrow g) = 0.0076 \,\,\,\text{W.u.}$ \cite{Minkov_Palffy_PRL_2017}), where  $n$ is the refractive index for the isomer photon, with the factor $n^3$  caused by enhancement of the $M1$ spontaneous decay in refractive media due to higher density of states of emitted photons \cite{Tkalya2000}.  We consider here $n^3=4$ for simplicity. Furthermore,  $\Delta$ is the detuning of the driving VUV laser to the nuclear transition energy, and $R$ the excitation rate. The latter can be expressed via the matrix elements $V_{m_1g_1}$ and $V_{m_2g_2}$ of the interaction Hamiltonian averaged over orientations of the electric field gradient as
\begin{equation}
R=\frac{\langle |V_{m_1g_1}|^2 + |V_{m_2g_2}|^2 \rangle}{3\gamma}=\frac{2 \pi}{15} \frac{c^2 I_0}{\hbar \omega_m^3}\frac{\Gamma}{\gamma}.
\label{eq:2}
\end{equation}
Here angular brackets denote averaging over spatial orientations of the electric field gradient, and $I_0$ is the intensity of the VUV clock driving radiation with frequency $\omega_m$. 

A single interrogation cycle consists of 4 time intervals: in the first and in the third of them (both have duration $\theta$) the sample is illuminated by the narrow-band VUV laser radiation whose frequency is detuned by $\delta_m$ to the blue and to the red side from the nominal frequency of the local oscillator respectively. The frequency offset $f$ (i.e., the difference between the nominal frequency of the local oscillator and the frequency of the isomer transition) is determined from the difference in the numbers $N_{n}$ and $N_{n+1}$ of photons counted during the second and the fourth time intervals (both have duration $\theta'$) respectively.

Mean numbers of photons counted in the second and fourth intervals can be expressed as
\begin{align}
\overline{N_{n}} &=a(f+\delta_m)+b(f+\delta_m)\overline{N_{n-1}},  \label{eq:3} \\
\overline{N_{n+1}}&=a(f-\delta_m)+b(f-\delta_m)\overline{N_{n}},  \label{eq:4}
\end{align}
where $\overline{N_{n-1}}$ is a mean number of photons measured in the fourth time interval of the previous interrogation cycle, and
\begin{align}
a(\Delta)&=\frac{2 \Neff}{5}\left(1-e^{-\Gamma^\zeta_{qu} \theta'} \right)
\left(1-\frac{\Gamma n^3}{G(\Delta)}\right)\left(1-e^{-G(\Delta) \theta} \right), \label{eq:5} \\
b(\Delta)&=e^{-G(\Delta) \theta-\Gamma^\zeta_{qu} \theta'}. \label{eq:6}
\end{align}
Here, $\Gamma^\zeta_{qu}$ is the EB decay rate of the isomer state in the presence of quenching optical laser field,  $\Neff=\NTh\cdot \frac{k\Omega}{4\pi}$ is the ``effective'' number of thorium nuclei ($k$ is quantum efficiency of the photodetector and $\Omega$ is the solid angle covered by this detector; we take $\Neff=10^{12}$, as in \cite{Kazakov_2012}), and
\begin{align}
G(\Delta)&=\Gamma n^3+\frac{5R/2}{1+\Delta^2/\gamma^2}. \label{eq:7}
\end{align}
For the relaxation rate of the nuclear transition coherences we use the value $\gamma=2 \pi \times 150~{\rm Hz}$ \cite{Kazakov_2012}.

If the offset $f$ of the local oscillator frequency from the clock transition frequency is small, we can express it as
\begin{equation}
f=\frac{\overline{N_{n+1}}-\overline{N_n}-b_0(\overline{N_n}-\overline{N_{n-1}})}{2a_1-b_1(\overline{N_n}-\overline{N_{n-1}})}, \label{eq:8}
\end{equation}
where $a(f \pm \delta_m)=a_0\mp a_1 f$; $b(f \pm \delta_m)=b_0\pm b_1 f$. 

Supposing that the numbers of photons $N_{n+1}$ (as well as $N_{n-1}$) and $N_{n}$ counted in the fourth and second time interval are Poissonian random numbers with means (\ref{eq:3}) and (\ref{eq:4}), one may estimate the error $\delta f$ of determination of the frequency offset $f$ as
\begin{equation}
\delta f=\frac{\sqrt{a_0(1-b_0)(1+b_0+b_0^2)}}{\sqrt{2}\left[a_1(1-b_0)-b_1a_0\right]}. \label{eq:9}
\end{equation}
This expression represents a fundamental lower limit of the error offset for a single interrogation cycle. 

\begin{figure}
\includegraphics[scale=1]{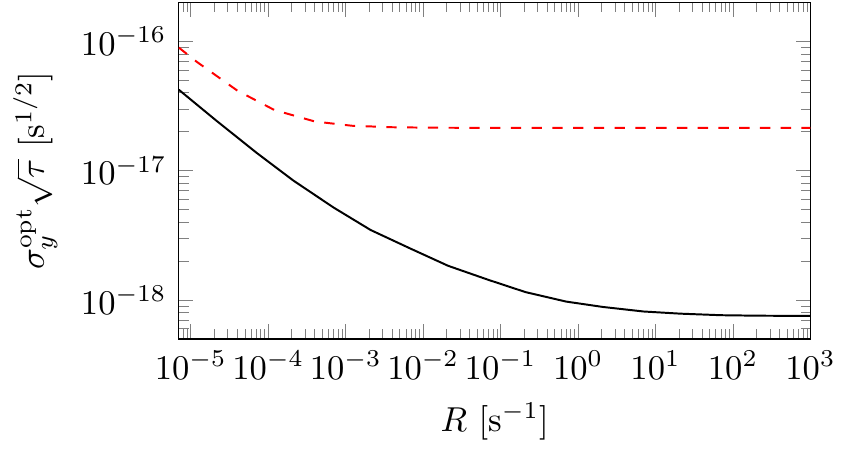}
\caption{Ultimate clock fractional instability $\sigma_y\sqrt{\tau}=\delta f \sqrt{t}/\omega$ as a function of excitation rate $R$ for an optimized interrogation cycle with (black solid curve) and without (red dotted curve) laser-enhanced quenching during measurement phases. See text for further explanations and used parameters. }
\label{fig:fClock2}
\end{figure}

To evaluate the possible improvement of the nuclear clock performance  that may be obtained with the help of laser-assisted quenching, one may consider the short-term instability defined as \cite{Kazakov_2012}  
\begin{equation}
\sigma_y(\tau)  = \frac{\delta f \sqrt{t}}{\omega \sqrt{\tau}}  \label{eq:10}
\end{equation}
where $t$ is the time of single interrogation cycle, and $\tau$ is the total measurement time. In order to reduce $\sigma_y$, one has to minimize $\delta f \sqrt{t}/\omega=\sigma_y\sqrt{\tau}$ by the proper choice of the intervals $\theta$ and $\theta'$ for the different phases of the interrogation cycle, and the working point $\delta_m$. Figure~\ref{fig:fClock2} presents an optimized $\sigma_y^{\rm opt}(\tau)\sqrt{\tau}$ as a function of the excitation rate $R$ which enters via Eqs.~\eqref{eq:5}, \eqref{eq:6} and \eqref{eq:7} the expression of $\delta f$. We compare the cases with and  without including the optical laser-driven quenching of the isomeric state during the measurement phases. For the latter case we replace the  EB quenching rate $\Gamma^\zeta_{qu}$ in Eqs.~\eqref{eq:5} and \eqref{eq:6} by the spontaneous radiative decay rate of the isomer. The parameters used in the calculations are $\Neff=10^{12}$, $\Gamma=10^{-4}~{\rm s^{-1}}$, $E_{m}=8.2~{\rm eV}$, and $\Gamma^\zeta_{qu}=0.07~{\rm s^{-1}}$. We suppose here that the local oscillator is perfectly stable, and the only detection noise is the shot noise of the detection of the isomer photons. The results in Fig.~{\ref{fig:fClock2}} show that  for strong enough excitation rates, the short-term stability may be improved by more than one order of magnitude using the quenching scheme. This makes the future experimental implementation of the quenching scheme very desirable. In order to achieve such high rates $R/\Gamma^\zeta_{qu} \ge10$, direct laser excitation of the isomer would require intensity $I_0=3.6$~W/cm$^2$ via Eq.~\eqref{eq:2}.

\section{Conclusion \label{concl}}

We have investigated driven EB processes in the  $^{229}$Th:CaF$_2$ solid state  environment  making use of defect states in the crystal electronic structure. These states are predicted by DFT within the crystal band gap, not far from the nuclear isomer energy, and would at first sight be considered a nuisance for  laser driving of the nuclear transition. Surprisingly, the defect states allow an efficient nuclear excitation via EB, up to two orders of magnitude stronger than photoexcitation. The rate of the EB excitation is dependent on the characteristics of the electronic defect states as well as the surrounding intermediate electronic states.  Questions still remain regarding the exact location of these defect states in energy, which we hope will be soon pinned down by experiments. Our calculations have mitigated this point by discussing a larger   resonance region  to illuminate how the system would change in the case of shifting electronic state energy. The nuclear transition  was shown to proceed upwards of $85\%$ via $E2$ multipolarity, while for the EB photon emission or absorption, the $E1$ bridge processes were dominant. Quenching of the isomeric state via the inverse bridge process was shown to significantly impact the potential stability of a solid state clock, with an increase by more than one order of magnitude.  Our theoretical models can be easily adjusted as more information regarding the crystal environment becomes known experimentally.

\section{Acknowledgements}

This work is part of the ThoriumNuclearClock project that has received funding from the European Research Council (ERC) under the European Union’s Horizon 2020 research and innovation programme (Grant agreement No. 856415). GK is supported by  the European Union’s Horizon 2020 Research and Innovation Programme No 820404 (iqClock project). AP gratefully acknowledges support from the Deutsche Forschungsgemeinschaft (DFG) in the framework of the Heisenberg Program. The computational results presented have been achieved in part using the Vienna Scientific Cluster (VSC). The authors also want to thank Peter Mohn for most valuable discussions.

\vfill
\newpage

\newpage	
\bibliographystyle{unsrt}
\bibliography{references}

\end{document}